\documentclass[twocolumn,twoside,slac_two]{revtex4}
\usepackage{graphicx}
\usepackage{fancyhdr}
\usepackage{revsymb}
\begin{document}

\title {Parsec-Scale Jet Behavior of the Quasar 3C~454.3 during the High Gamma-Ray
States in 2009 and 2010}

%

\author{S.G. Jorstad, A.P. Marscher, M. Joshi, N.R. MacDonald, T.L. Scott, and K.E. Williamson}
\affiliation{Institute for Astrophysical Research, Boston University, Boston, MA 02215, USA}
\author{P.S. Smith}
\affiliation{Steward Observatory, Tucson, AZ 85721, USA}
\author{V.M. Larionov}
\affiliation{St. Petersburg State University, St.Petersburg, 198504, Russia}
\author{I. Agudo}
\affiliation{Instituto de Astrof\'{\i}sica de Andaluc\'{\i}a (CSIC), Granada 18080, Spain \& IAR,
Boston University, Boston, MA 02215, USA}
\author{M. Gurwell}
\affiliation{Center for Astrophysics, Cambridge, MA 02138, USA}

\begin{abstract}
We analyze total and polarized intensity images of the quasar 3C~454.3
obtained monthly with the VLBA at 43 GHz  within the ongoing
Boston U. monitoring program of $\gamma$-ray blazars started in June 2007. The data
are supplemented by VLBA observations performed during intense campaigns of 2 week duration
when the quasar was observed 3 times per campaign. We find a strong increase of activity in
the parsec-scale jet of the quasar during high $\gamma$-ray states in December
2009, April 2010, and November 2010. We detect new superluminal knots, K09 and K10, associated with 
the autumn 2009 and 2010 outbursts, respectively, and compare their kinematic parameters. 
We analyze optical polarimetric behavior along with polarization parameters of the parsec-scale jet 
and outline similarities and differences in polarization properties across wavelengths.
The results of the analysis support the conclusions that the optical polarized emission is produced
in a region located in the vicinity of the mm-wave core of the jet of the quasar, and that
the $\gamma$-ray outbursts occur when a superluminal disturbance passes through the core.  
\end{abstract}

\maketitle

\thispagestyle{fancy}


\section{INTRODUCTION}
\begin{figure*}[tpb]
\centering
\includegraphics[width=135mm]{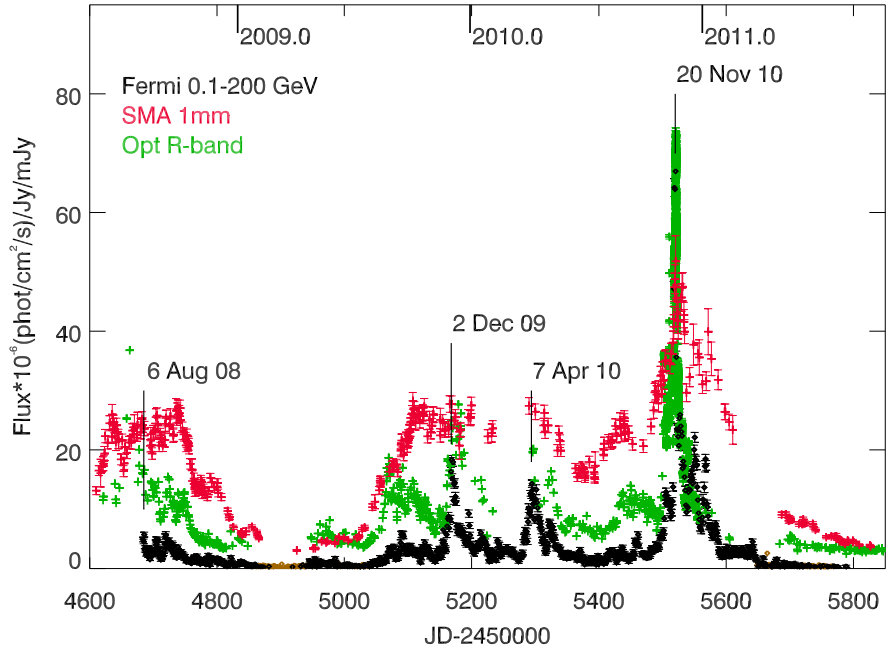}
\caption{Gamma-ray light curve of the quasar 3C~454.3, for energies E$>$100~MeV, with a 1-day binning interval (black, with orange points indicating upper limits) superposed on optical flux measurements in R band in mJy (multiplied by a factor of 3 for clearity; green), and the light curve at 225~GHz (1.3 mm, in Jy; red).}
 \label{glc}
\end{figure*}

The quasar 3C~454.3 (z=0.859) was extremely active at $\gamma$-ray energies
during the three years after the {\it Fermi} Large Area Telescope (LAT) started 
to operate in 2008 Summer. Figure~\ref{glc} shows the $\gamma$-ray light curve, for energies E$>$100~MeV,
of the quasar that we have constructed using {\it PASS 7 Photon and Spacecraft} data and 
the {\it Fermi}-LAT ScienceTools software package (ver. v9r23p1) with the 
P7\_V6\_DIFFUSE set of instrument response functions. Figure~\ref{glc} also displays the mm-wave (225~GHz)
light curve obtained at the Submillimeter Array (Mauna Kea, Hawaii), as well as the optical light curve
in R-band compiled from data from a number of ground-based telescopes. Four major peaks of the $\gamma$-ray
emission are seen in Figure~\ref{glc}, corresponding to (I) the 2008 July-August enhanced activity with an average flux 
$\sim$3$\times$10$^{-6}$~ph~cm$^{-2}$s$^{-1}$ \citep{Abdo09}; (II) the prominent $\gamma$-ray outburst in November-December 
2009 when the $\gamma$-ray emission reached F$_{100}$=22$\pm$1$\times$10$^{-6}$~ph~cm$^{-2}$s$^{-1}$ \citep{Acker10}; (III) 
the strong  $\gamma$-ray outburst in April 2010 with F$_{100}$=16$\pm$3$\times$10$^{-6}$~ph~cm$^{-2}$s$^{-1}$; and 
(IV) the gigantic outburst in November-December 2010 with a 3~hr peak F$_{100}$=85$\pm$5$\times$10$^{-6}$~ph~cm$^{-2}$s$^{-1}$ \citep{Abdo11}, which correponds to an apparent $\gamma$-ray luminosity of L$_\gamma$=2.1$\pm$0.2$\times$10$^{50}$~erg~s$^{-1}$ (H$_\circ$=71~km~s$^{-1}$, $\Omega_{m}$=0.27, and $\Omega_\lambda$=0.73). 

The $\gamma$-ray spectrum of the quasar is well described
by a broken power law with photon index $\Gamma\sim$2.3 below the break frequency $\sim$2~GeV and $\Gamma\sim$3.5 above the break
\citep{Abdo09}. During outbursts I-III the $\gamma$-ray spectrum experienced a very moderate hardening when the $\gamma$-ray flux increased \citep{Acker10}. However, during the last outburst the spectrum showed significant differences between light curves for fluxes F$_{0.1-1GeV}$ and F$_{>1GeV}$, with a progressive decrease of photon index from $\Gamma\approx$2.35 to $\Gamma\approx$2.1 as the most prominent flares developed.  Comparison of the flux measured at 31~Gev with the flux extrapolated from lower energy suggests an upper limit of the photon-photon optical depth
$\tau_{\gamma\gamma}\approx$2, which implies that
the emission region is located at $z_{em}\approx$0.14~pc from the central engine, which is close to, or beyond, the expected outer boundary of the BLR \citep{Abdo11}. The short timescale of flux variability (by a factor of 2 over 3 hr) is considered as indication 
of the origin of the $\gamma$-ray emission region within the BLR\citep{TAV10}. However, a timescale of the variability constrains the size of the flaring region, not its location. 

The multi-frequency behavior of the quasar is complex. There is a strong correlation between long-term variations at optical and $\gamma$-ray frequencies, although
the correlation weakens at short timescales \citep{J10,RAI11,VER11}. Raiteri et al. \citep{RAI11} note that during outbursts I-III the same maximum optical flux was reached and the peaks at mm-waves are also comparable, while the $\gamma$-ray flux was significantly higher during outbursts II \& III with respect to I. The authors interpret the optical through mm-wave observations as synchrotron radiation produced in an inhomogeneous jet with a progressive increase of both the size and distance (from the jet apex) of emitting region
with wavelength. In this scenario the $\gamma$-ray emission is produced slightly downstream with respect to
the optical synchrotron emission region if the synchrotron-self-compton mechanism (SSC) is responsible
for $\gamma$-ray production. The $\gamma$-ray outburst IV has prominent counterparts at longer wavelengths (from X-ray
to mm-wave) and the peaks of the outburst coincide within 1~day at different frequencies (see Fig.~\ref{glc}). However, the amplitude
of $\gamma$-ray variations is significatly higher than those at longer wavelengths. Vercellone et al. \citep{VER11} have modeled such behavior by an external Compton component boosted by a local enhancement of soft seed photons.

We monitor 3C~454.3 with the VLBA 
at 43~GHz in total and polarized intensity and at optical wavelengths in photometric and polarimetric modes. In this paper
we add analysis of the parsec-scale jet and polarization  behavior during the outbursts II-IV to the multi-frequency studies of the blazar, and thereby tighten constraints on the location of the $\gamma$-ray production.

\section{Parsec-Scale Jet Kinematics}
\begin{figure*}[tpb]
\centering
\includegraphics[width=105mm]{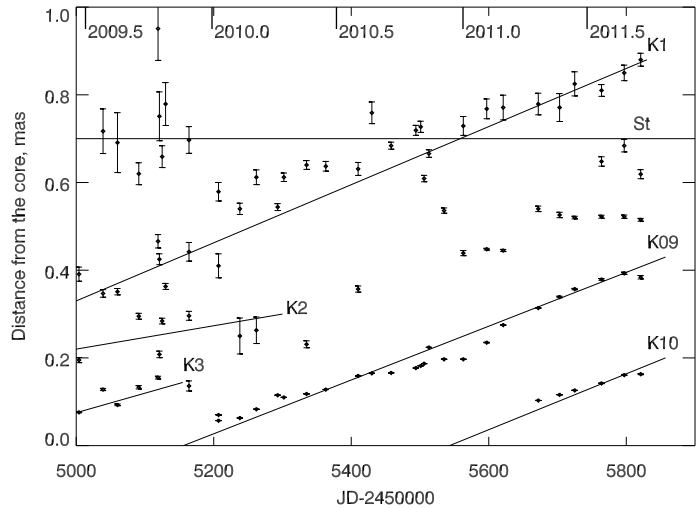}
\caption{Position of jet features with respect to the core as a function of time.}
 \label{evol}
\end{figure*}

\begin{figure*}[tpb]
\centering
\includegraphics[width=105mm]{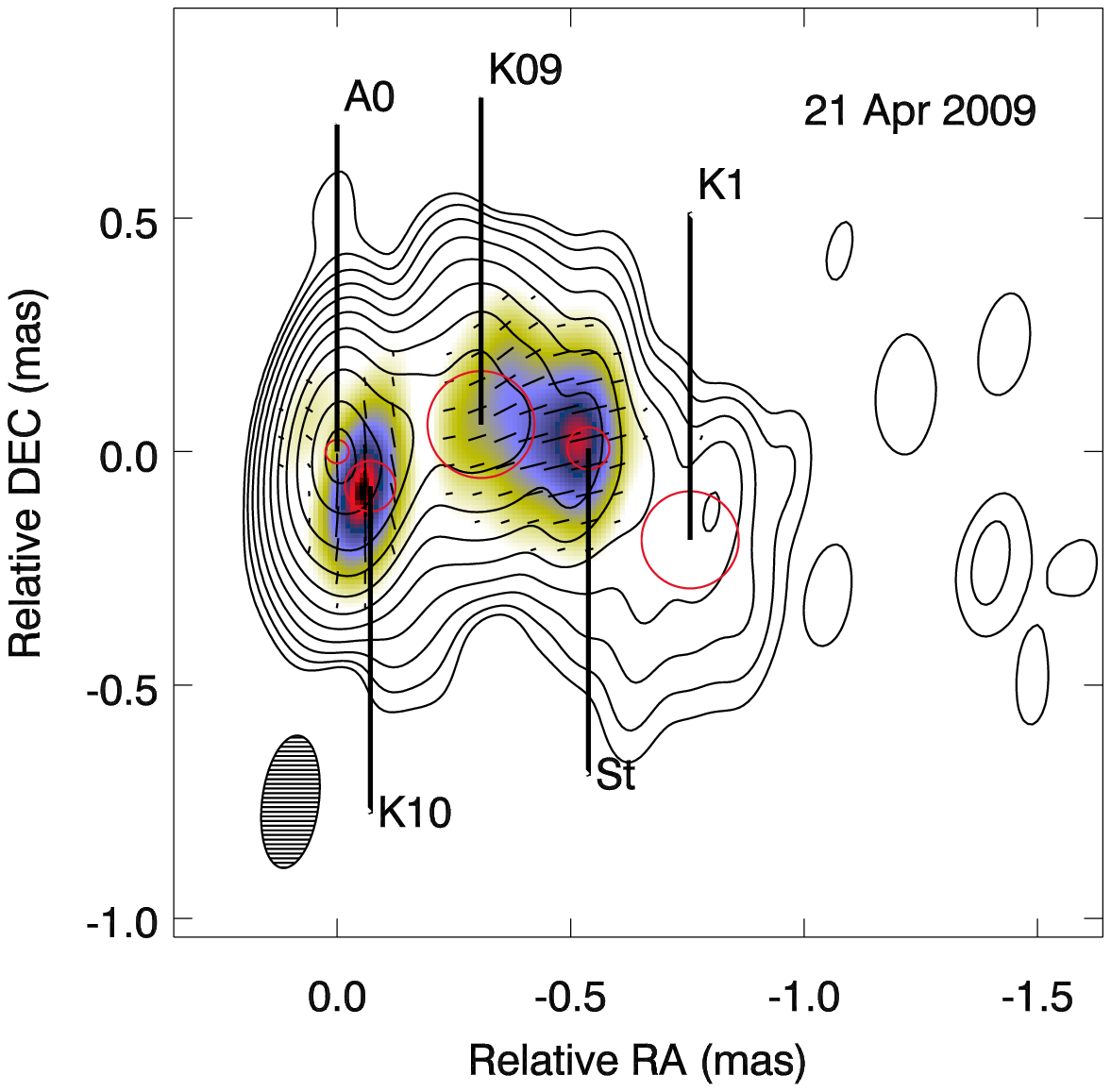}
\caption{Total (contours) and polarized (color scale) intensity images of 3C~454.3, with a total intensity peak of 7.62~Jy/beam,
polarized intensity peak of 263~mJy/beam, and beam = 0.33$\times$0.14~mas$^2$ at PA=$-10^\circ$. Contours start at 0.125\%
of the peak and increase by a factor of 2. Red circles indicate positions and sizes of components according to the model fitting. Black line segments within
the image indicate direction of linear polarization.}
 \label{map}
\end{figure*}

We monitor 3C~454.3 roughly monthly with the VLBA at 43 GHz during 24~hr session of observations of 30 $\gamma$-ray blazars. These
observations are supplemented by observations carried out during four campaigns of two-week duration each (in 2008 October, 2009 October, 2010 April, and 2010 November) when 12-15 sources were observed during each 16~hr session three
times per campaign. This resulted in 49 epochs of VLBA observations of the quasar obtained from 2008 January to 2011 September. The VLBA data were processed and imaged in a manner identical to that described in Jorstad et al. \citep{J05}. We performed calibration of the position angle of polarization (EVPA) using (1) VLA observations carried out during the campaigns, coinciding with one of the VLBA epochs, (2) the {\it D-term}  
method \citep{Dterm}, and (3) the stability of the EVPAs of selected stationary features in sources observed along with 3C~454.3. In general, the accuracy of the EVPA calibration is within 5-10 degrees. We model the images in terms of a small number of components with circular Gaussian brightness distributions and determine polarization parameters of components
using an IDL program that calculates the mean values of the pixels at the position of a total intensity component and within an area equal to that of the size established by the model fit. Figure~\ref{evol} shows the results of modeling within 1 mas
of the core of 3C~454.3, which is a presumably stationary feature located at the eastern end of the jet. In addition to the core, A0,
three features are prominent in the jet: a quasi-stationary component, St, which was seen previously (e.g., \citep{J05})
and two new suprluminal knots, K09 and K10. Three other features in the jets, K1, K2, and K3, appear to correspond to the knots discussed in \citep{J10}. They are designated with respect to those identifications. Figure~\ref{map} shows the total and polarized image on 2011 April 21, where four components (A0, St, K09, and K10) are indicated according to their parameters found by model fitting. 
The kinematics of both knots K09 and K10 is complex. Knot K09 does not move ballistically, as is seen in Figure~\ref{evol}.
It decelerates at a distance of $\sim$0.15~mas from the core and then accelerates at a distance of $\sim$0.3~mas.
Knot K10 appears in the jet at PA$\approx-136^\circ$, to the south with respect to the jet axis (Fig.~\ref{map}), then its trajectory curves into the direction of the jet axis, PA$\approx-98^\circ$. Table~\ref{Kparm} lists parameters of the knot's kinematics: proper motion, $\mu$; angular acceleration along the jet, $\dot{\mu}_\parallel$; angular acceleration perpendicular to the jet, $\dot{\mu}_\perp$; apparent speed, $\beta_{\rm app}$;
time of ejection, $T_\circ$; maximum flux, $S_{\rm max}$; timescale of flux variability, $t_{\rm var}$; and 
angular size of component at the epoch of maximum flux, $a$. Table~\ref{Kparm} also gives the Doppler factor, 
$\delta$, bulk Lorentz factor, $\Gamma_b$, viewing angle, $\Theta_\circ$, and Doppler factor, $\delta$, based on the approach that we developed in Jorstad et al. \citep{J05}. The parameters $\delta$, $\Gamma_b$, and $\Theta_\circ$ of K09 are very close to those which we derived
using the kinematics of the jet in 1998-2001 \citep{J05}, while the parameters of K10 agree with those of knot K3 \citep{J10}, which implies a Doppler factor of the jet of $\sim$50 during the most prominent $\gamma$-ray outburst. In addition, strong intraday variability is seen in R band during 2010 November 18-21, which coincides with the peak of the $\gamma$-ray outburst and the emergence of K10 from the core. Intraday variability was observed also during the optical outburst at the end of 2007, which we associated with knot K3 \citep{J10}. The apparent speeds of K09 and K10 are similar and higher than $\beta_{app}$ of K1, K2, and K3 reported in \citep{J10},
although within the range of $\beta_{app}$ found in \citep{J05}. According to the analysis, the difference in the derived value of $\delta$ of K09 and K10 results from faster fading of K10 on the VLBA images with respect to K09 (see $t_{\rm var}$ in Table~\ref{Kparm}). 

\begin{table}[t]
\begin{center}
\caption{Kinematics of Superluminal Knots}
\begin{tabular}{|l|c|c|}
\hline \textbf{Parameter} & \textbf{K09} & \textbf{K10}
\\
\hline
$\mu$, mas~yr$^{-1}$&0.21$\pm$0.02&0.19$\pm$0.03 \\
$\dot{\mu}_\parallel$, mas~yr$^{-2}$&0.10$\pm$0.01&-- \\
$\dot{\mu}_\perp$, mas~yr$^{-2}$&0.13$\pm$0.02&0.41$\pm$0.22 \\
$\beta_{\rm app}$, $c$&9.6$\pm$0.6&8.9$\pm$1.7 \\
$T_\circ$, yr &2009.88$\pm$0.05&2010.95$\pm$0.07 \\
$T_\circ$, RJD &5156$\pm$18&5543$\pm$25 \\
$S_{\rm max}$, Jy&17.00$\pm$0.45&7.10$\pm$0.16 \\
$t_{\rm var}$, yr&0.67$\pm$0.06&0.26$\pm$0.02 \\
$a$, mas&0.12$\pm$0.02&0.10$\pm$0.01 \\
$\delta$&27$\pm$3&58$\pm$6 \\
$\Gamma_b$&15$\pm$2&29$\pm$3 \\
$\Theta_\circ$, deg&1.35$\pm$0.2&0.3$\pm$0.1 \\
$N$&24& 6 \\
\hline
\end{tabular}
\label{Kparm}
\end{center}
\end{table}
\begin{figure*}[tpb]
\centering
\includegraphics[width=135mm]{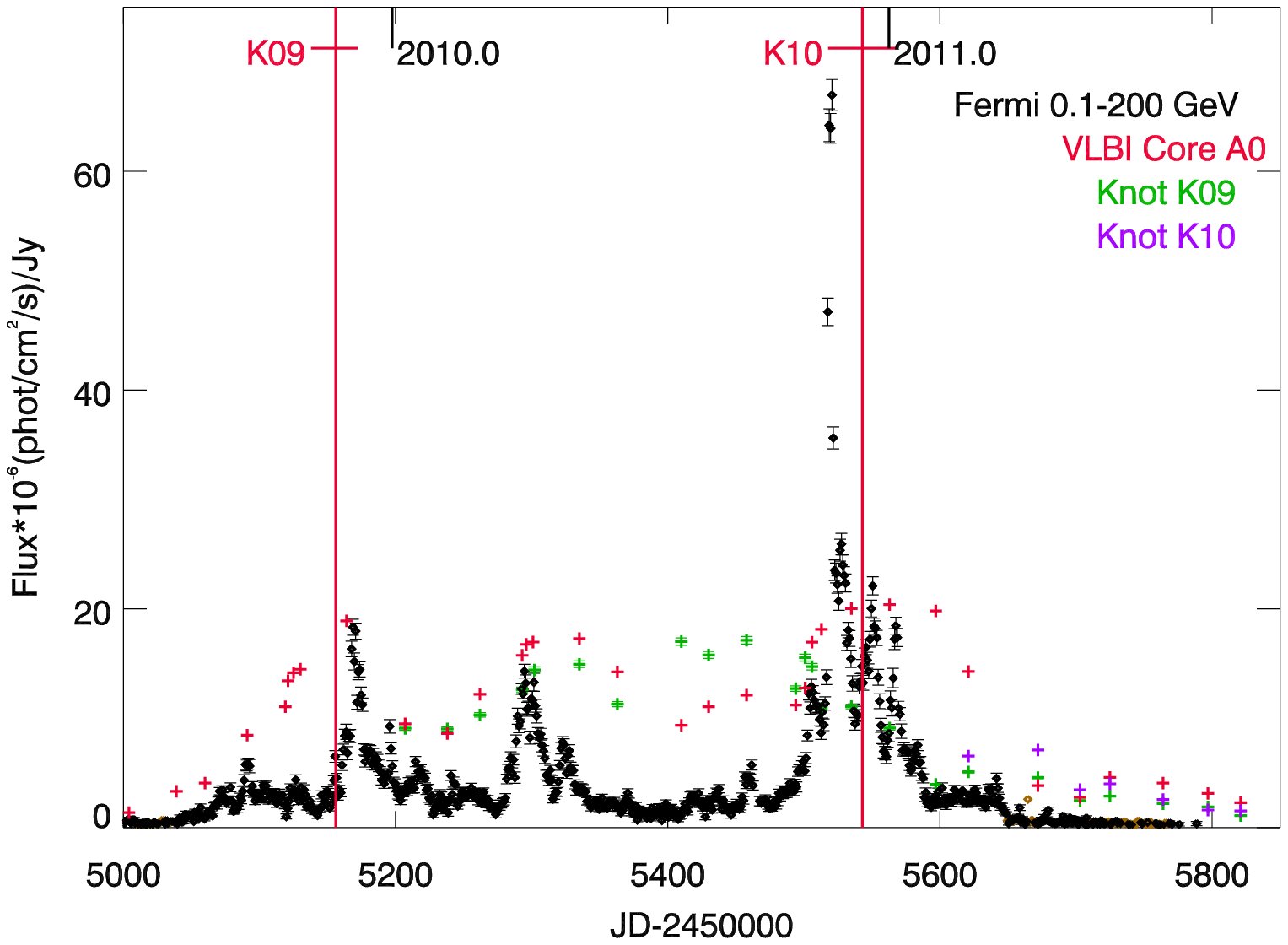}
\caption{Gamma-ray (black), VLBI core at 7 mm (red), K09 (green), and  K10 (violet) light curves;
red lines show times of passage of knots K09 and K10 through the VLBI core.}
 \label{gvlc}
\end{figure*}

\begin{figure*}[tpb]
\centering
\includegraphics[width=95mm]{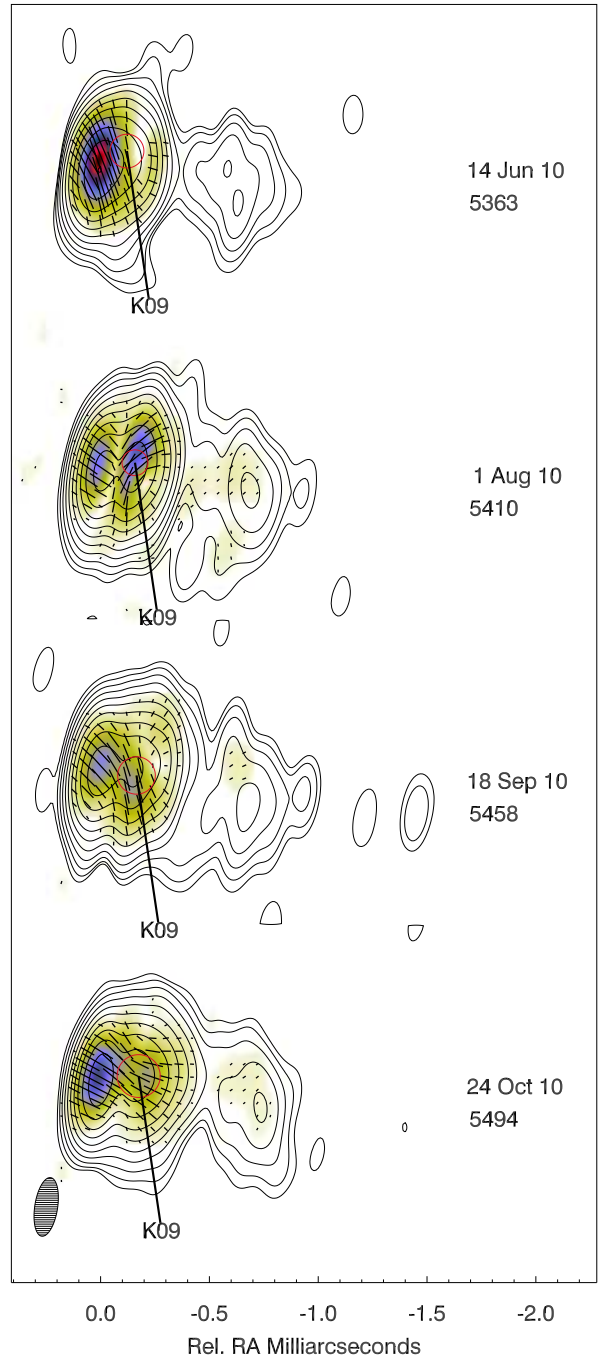}
\caption{The total (contours) and polarized (color scale) intensity images of 3C~454.4 with the total intensity peak 13.66~Jy/beam
and polarized intensity peak 633~mJy/beam and beam=0.33$\times$0.14~mas$^2$ at PA=-10$^\circ$. Contours start at 0.125\%
of the peak and increase by a factor of 2. The red circles indicate position and size of knot K09 according to model fitting; the black line segments within
each image indicate direction of linear polarization.}
 \label{map4mod}
\end{figure*}

Figure~\ref{gvlc}
plots the $\gamma$-ray light curve and light curves of the core, A0, and components K09 and K10, and indicates
the times of ejection of the knots.  
It shows that disturbances K09 and K10 passed through the mm-wave VLBI core
close (within the 1$\sigma$ uncertainty of the ejection time of 18 and 25~days, respectively) to the time of $\gamma$-ray peaks of the prominent outbursts in autumn 2009 and 2010. 
Figure \ref{gvlc} also reveals that during the strong $\gamma$-ray outburst in 2010 April both the core and  K09 had an outburst as well. Therefore, based on the history of 3C~454.3, it is likely that the increase in the flux of the core 
in 2010 April during the $\gamma$-ray event was caused by passage of a new superluminal knot through the core, with a flux of $\le$5-7~Jy. We note that both the core was very bright in 2010 April  (S$\sim$18~Jy) and K09 was of similar brightness up to end of 2010. Summer - Autumn 2010 would have been the best period for detection of this hypothetical new component in the jet.
Over this period K09 had a relatively small separation from the core, $\sim$0.12-0.20~mas. Although K09 was well resolved with the high resolution of our observations ($\sim$0.1~mas, see Fig.~\ref{map4mod}), the resolution is most likely insufficient to have resolved a structure consisting of a weaker knot between two extremely bright ones within 0.20~mas. The increase in the flux of K09 in 2010 April might be an artifact of modeling by circular gaussians of such complex structure, but K09 continued to be
bright (brighter than the core) when the core faded in the middle of 2010 (Fig.~\ref{gvlc}) that suggests that K09 (perhaps, blended with another superluminal knot) was responsible for the plateau at 1~mm 
(Fig.~\ref{glc}) on which the Autumn 2010 outburst is superposed. The new dramatic increase in the core flux in November 2010 and ejection of K10 further reduced the possibility of detecting any propagating disturbance associated with the $\gamma$-ray event in April 2010. However, collection of these events during the time of the high $\gamma$-ray state is strong evidence of a tight connection between activity in the parsec-scale jet and enhanced $\gamma$-ray emission. 
 
\section{Polarization Behavior}
\begin{figure*}[tpb]
\centering
\includegraphics[width=135mm]{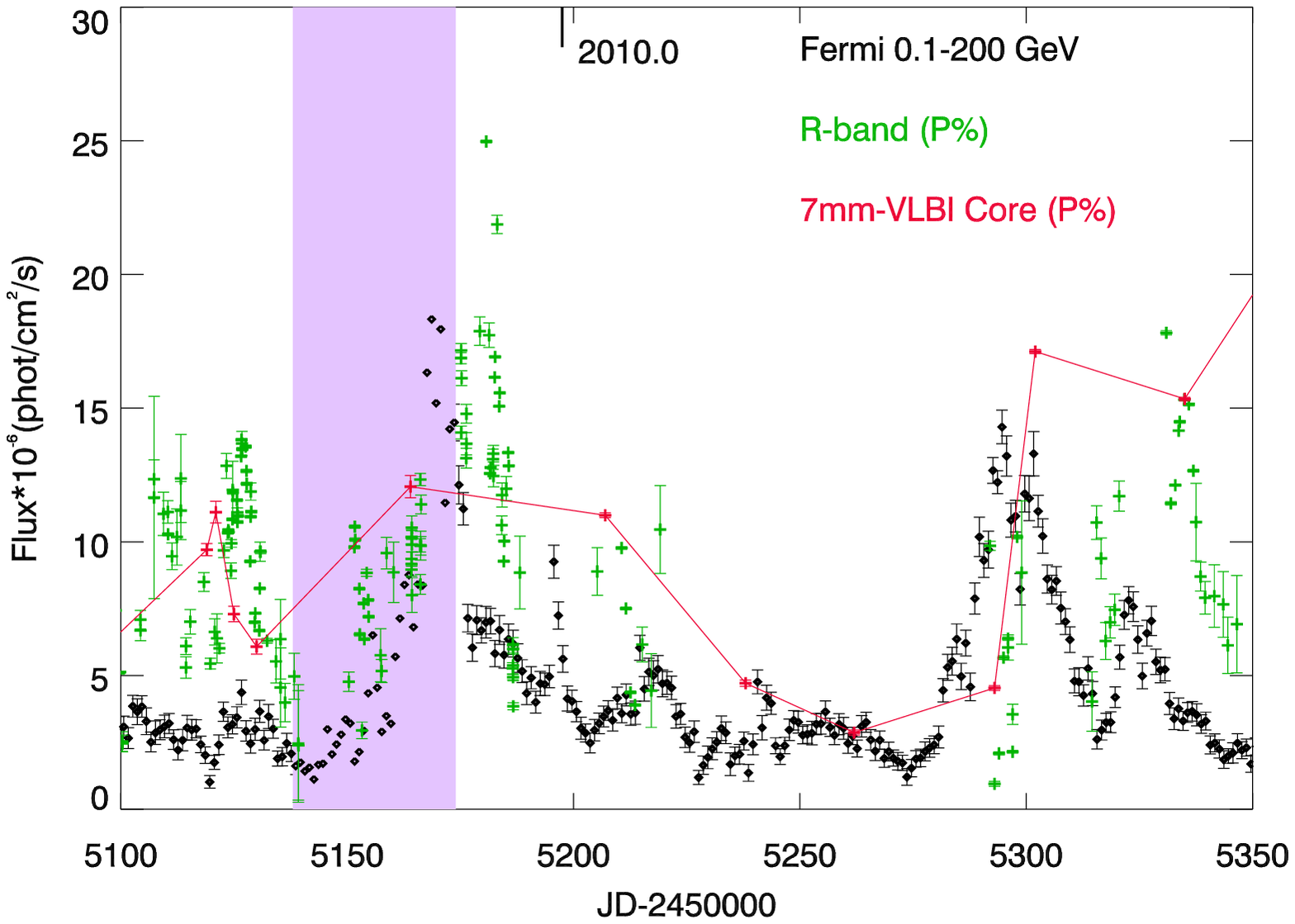}
\caption{Gamma-ray (black) and degree of polarization light curves at optical wavelengths (green) and in the VLBI core at 7 mm (red) during the autumn 2009 and Spring 2010 $\gamma$-ray outbursts; degree of polarization at 7 mm is multiplied by a factor of 5. The violet area indicates the time of passage of knot K09 through the VLBI core.}
 \label{pol1}
\end{figure*}

\begin{figure*}[tpb]
\centering
\includegraphics[width=135mm]{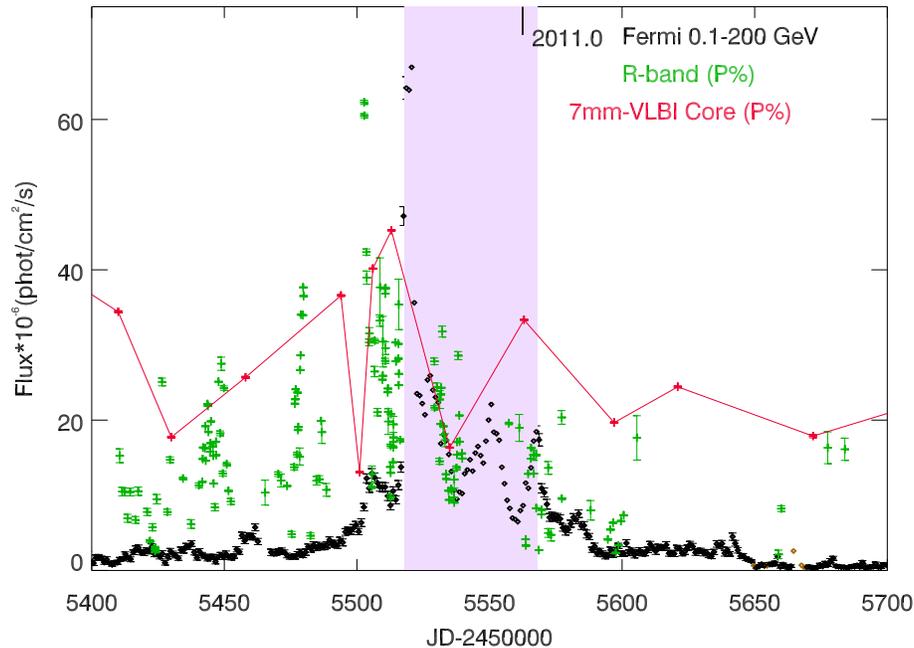}
\caption{Gamma-ray (black) and degree of polarization light curves at optical wavelengths (green) and in the VLBI core at 7 mm (red) during the autumn 2010 $\gamma$-ray outburst; degrees of optical polarization and at 7 mm are multiplied by factor of 2 and 10, respectively.  Violet area indicates the time of passage of knot K10 through the VLBI core.}
 \label{pol2}
\end{figure*}

An increase of the degree of optical linear polarization, $p$, and rotation of the polarization position angle, $\chi_{opt}$, during high $\gamma$-ray/optical states were observed previously
in several sources \citep{MAR10,Abdo10}, as well as in 3C~454.3 \citep{J10,SAS11}. Sasada et al. \citep{SAS11}
report two different episodes of rotation of $\chi_{opt}$ during the Autumn 2009 event - a fast rotation of $-26.3\pm$2.3~deg/day (clockwise) in the beginning of the active state and a slower rotation of 9.8$\pm$0.5~deg/day (counterclockwise) from RJD:5157 to RJD:5192. The latter coincides with the peak of the outburst and passage of knot K09 through the VLBI core. A strong correlation between the gamma-ray flux and optical polarization at the early stage of the 2009 outburst was reported in Smith et al.\citep{SMITH09}, when the degree of the polarization decreased from $p\sim$12\% to 3\% within a day in unison with the $\gamma$-ray flux. Our polarimetric monitoring during $\gamma$-ray events II, III, and IV shows that
changes of $p$ in the 7~mm core are similar to the general behavior of the optical polarization (Figs.~\ref{pol1},\ref{pol2}).  This supports the idea that the polarized optical emisssion is produced in the vicinity of the mm-wave core of the blazar's jet, as previously proposed by our group\citep{J07,Frani07,IVAN1,IVAN2}. Moreover, 
a strong increase in the degree of polarization both at optical wavelengths and in the VLBI core occurs 
at or near the $\gamma$-ray peaks. This implies that ordering of the magnetic field is needed for high $\gamma$-ray production in blazars. The most plausible mechanism for ordering of the magnetic field is the presence of a shock. Hughes, Aller, \& Aller \citep{HAA11} find that oblique shocks can reproduce variations in centimeter-wave
polarization observed during $\gamma$-ray outbursts in blazars. However, the optical degree of polarization
rises up to 25-30\% (perhaps even higher, since in 2010 November our polarimetric observations missed the peak of
the $\gamma$-ray emission), while the maximum of $p$ in the 7~mm core is only $\sim$5\%. The latter implies that the polarized optical emission region, with a nearly uniform magnetic field, occupies only $\sim$0.04 of the mm-wave core. This suggests that the rest of the core contains a fairly turbulent magnetic field. Such a situation can be accommodated in the turbulent extreme multi-zone model proposed by Marscher~\citep{MAR12}, in which a chaotic magnetic field modulates the efficiency of particle acceleration as the jet plasma crosses shock fronts.    
\begin{figure*}[tpb]
\centering
\includegraphics[width=135mm]{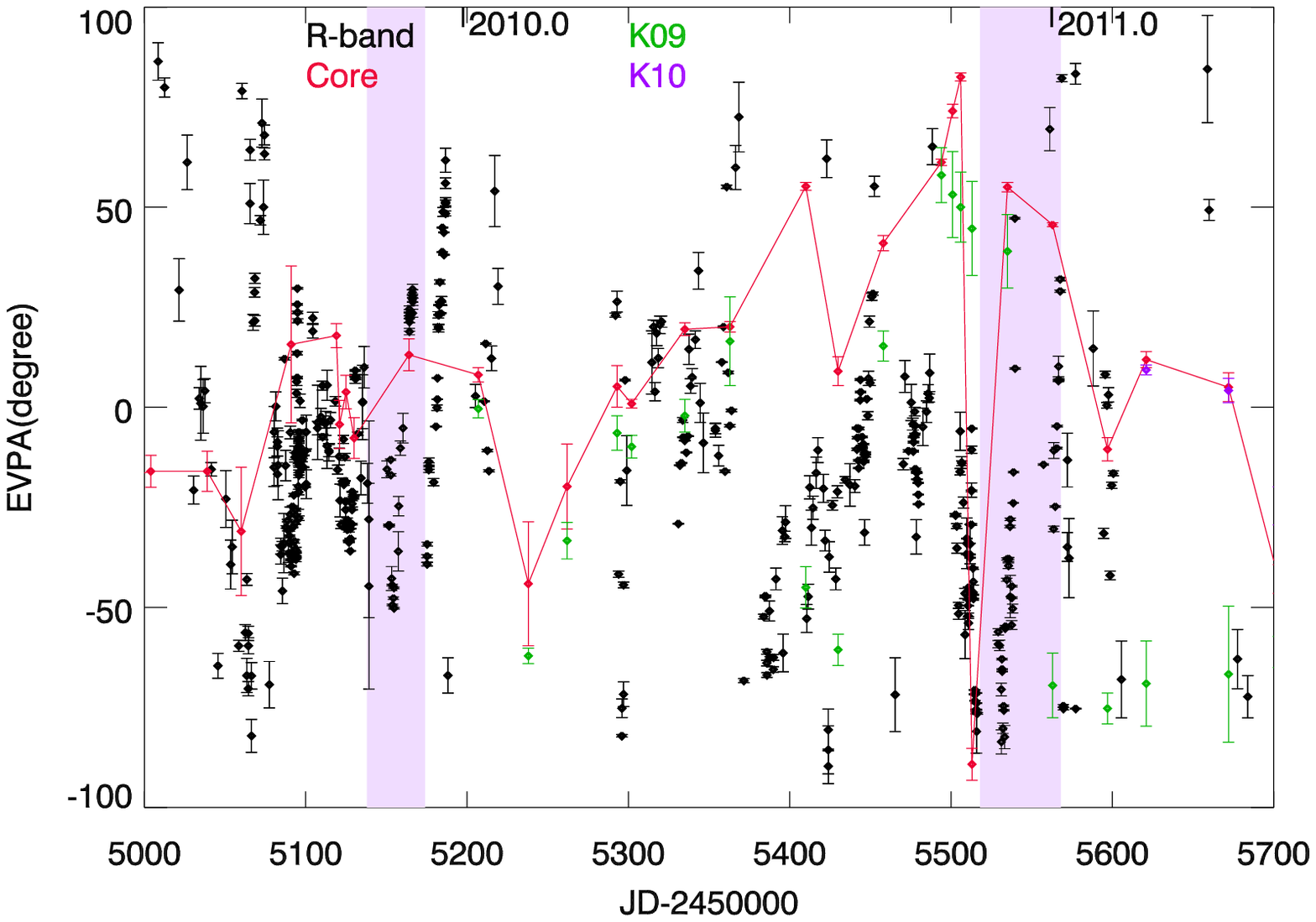}
\caption{Position angle of polarization at optical wavelengths (black), in the VLBI core at 7 mm (red),
in knot K09 (green), and in knot K10 (violet). Violet areas indicate the times of passage of K09 and K10 through the VLBI core.}
 \label{evpa}
\end{figure*} 

Figure~\ref{evpa} shows the behavior of the position angle of polarization at optical wavelengths, $\chi_{opt}$, and in the jet at 7 mm, $\chi_{core}$, $\chi_{K09}$, and $\chi_{K10}$. In general, $\chi_{opt}$ varies
over the whole interval of possible position angles, while $\chi_{core}$ tends to be confined within
position angles of -20-+10$^\circ$ (57\% of the measurements), which corresponds to a direction of the polarization that is nearly perpendicular to the jet axis, $\Theta\sim -98^\circ$.
Figure~\ref{evpa} reveals a rotation of $\chi_{opt}$ during passage of both knots K09 (the event
observed by Sasada et al.\citep{SAS11}) and K10 through the core.
Moreover, there is a good agreement between $\chi_{opt}$ and $\chi_{core}$ measured during the Autumn 2009 outburst, and there is a signature of a rotation of  
$\chi_{core}$ in a similar manner as $\chi_{opt}$ does during the Autumn 2010 outburst, althouth the 7~mm data are sparse. These are very strong indications that the optical polarized emission region is tightly connected to the mm-wave core of the radio jet. 

 \section{Conclusions}
Multifrequency observations of the quasar 3C~454.3 show a strong connection between the $\gamma$-ray, optical, 1~mm, and 7~mm VLBI core light curves. The well-defined peaks of the optical and 1~mm light curves
coincided within 1~day with the peak of the gigantic $\gamma$-ray outburst in 2010 November. This argues 
in favor of the conclusion that the $\gamma$-ray and optical emission regions are co-spatial with that at millimeter wavelengths. During two major $\gamma$-outbursts, in 2009 December and 2010 November, a superluminal knot emerged from the core at the time of the $\gamma$-ray event (within the uncertainties of the VLBI measurements). The degrees of optical and core polarization reached maxima near to or at the same time as the
$\gamma$-ray peaks (depending on the data sampling). Furthermore, the position angle of optical polarization rotated when a superluminal knot emerged from the core. Although we did not detect a disturbance associated with the $\gamma$-ray event
in 2010 April, an increase in the flux and polarization in the core region was observed for this event as well, suggesting that such a disturbance was indeed passing through the core at this time.
These observational findings strongly support the idea that the $\gamma$-ray outbursts originate in the mm-wave 
core of the jet when a disturbance, propagating down the jet at a relativistic speed, passes through the core. According to Jorstad et al.~\citep{J05}, the parsec-scale jet of the quasar has a half opening angle $\sim$0.8$^\circ$. During the time period discussed here, the average angular size of the core at 43~GHz was $\sim$0.06~mas (although the core
is not fully resolved in the images). This locates the region of the high energy activity in the quasar 3C~454.3 at a distance $\le$16~pc from the central engine. 

\begin{acknowledgments}
This research was supported in part by NASA grants NNX08AV65G, NNX08AV61G,
NNX09AT99G, NNX10AU15G, and NNX11AQ03G, and NSF grant AST-0907893 (Boston U.), NASA grants NNX08AW56G and NNX09AU10G (Steward Obs.).  Agudo acknowledges funding by the ``Consejer\'{\i}a de Econom\'{\i}a, Innovaci\'on 
y Ciencia'' of the Regional Government of Andaluc\'{\i}a through grant P09-FQM-4784, and by the ``Ministerio
de Econom\'{\i}a y Competitividad'' of Spain through grant AYA2010-14844. The VLBA is an
instrument of the National Radio Astronomy Observatory, a facility of the Nation
al Science Foundation operated under cooperative agreement by Associated Universities, Inc.
 
\end{acknowledgments}

\bigskip 

\end{document}